\documentclass[12pt]{article}

\usepackage{graphicx}
\usepackage{cite}
\usepackage{times}

\usepackage{color}

\newcounter{lastnote}

\title{\textbf{\huge
The maintenance of sex in bacteria is ensured by its potential to reload
genes
}}

\author{Gergely J. Sz\"oll\H{o}si,$^\ast$ Imre Der\'enyi,$^{\ast}$  Tibor
  Vellai$^{\dagger}$\\
\\
\normalsize{$^{\ast}$Department of Biological Physics, E\"otv\"os University,}\\
\normalsize{P\'azm\'any P.\ stny.\ 1A,}
\normalsize{H-1117 Budapest, Hungary}\\
\normalsize{$^{\dagger}$Department of Genetics, E\"otv\"os University,
  Budapest, Hungary}\\
\normalsize{P\'azm\'any P.\ stny.\ 1C,}
\normalsize{H-1117 Budapest, Hungary}\\
\\
\\
\\
\\
}
\date{}

\begin{document}
\maketitle

\newpage
\begin{center}
\normalsize{\bf Running head:} \\
\normalsize{Sex is maintained by gene reloading}
\vskip 0.5 cm
\normalsize{\bf Keywords:}\\ 
\normalsize{natural genetic transformation, bacterial competence,}\\
\normalsize{evolution of sex, genome organization in prokaryotes}
\vskip 0.5 cm
\normalsize{\large \bf To whom correspondence should be addressed:} \\
\vskip 0.15 cm 
\normalsize{\large Imre Der\'enyi }\\
\normalsize{Department of Biological Physics, E\"otv\"os University,}\\
\normalsize{P\'azm\'any P.\ stny.\ 1A, H-1117 Budapest, Hungary}\\
\normalsize{E-mail: derenyi@angel.elte.hu,}\\
\normalsize{Tel: (+36-1) 372-2795,} \\
\normalsize{Fax: (+36-1) 372-2757 }\\
\end{center}
{
\begin{center}
{\large ABSTRACT}
\end{center}

Why sex is maintained in nature is a fundamental question in
biology. Natural genetic transformation (NGT) is
a sexual process by which bacteria actively take up exogenous DNA and
use it to replace homologous chromosomal sequences. As it has been
demonstrated, the role of NGT in repairing deleterious mutations under constant
selection is insufficient for its survival, and the lack
of other viable explanations have left no alternative except that DNA uptake
provides nucleotides for food. 
Here we
develop a novel simulation approach for the long-term dynamics of
genome organization (involving the loss and acquisition of genes) in a bacterial species consisting of
a large number of spatially distinct
populations subject to independently fluctuating ecological
conditions. Our results show that in the presence of weak
inter-population migration NGT is able to subsist as a mechanism to reload
locally lost, intermittently selected genes from the collective gene
pool of the species through DNA uptake from migrants. Reloading genes
and 
combining them with those in locally adapted genomes allow individual cells to
re-adapt faster to environmental changes. 
The machinery of transformation survives under a wide range of model
parameters readily encompassing real-world biological conditions. 
These findings imply that the primary role of NGT is not to serve the cell with food,
but to provide homologous sequences for restoring genes 
that have disappeared from or become degraded in the local population.
}

\newpage

\begin{center}
{\large INTRODUCTION}
\end{center}
Sexual reproduction is a process that brings genomes, or portions of
genomes, from different individuals into a common cell, producing a
new combination of genes: in eukaryotes, this occurs as a result of
fertilization and meiotic recombination; in bacteria, it happens as a
result of the acquisition of exogenous DNA.  The ubiquity of genetic
transfer in bacteria is reflected in the  dynamic structure of their genomes,
which are constantly being shaped by two opposing forces: selection
for shorter length (favoring DNA loss through deletion) and selection
for gene function (driving genome loading by the acquisition of
exogenous DNA) (\citen{Vellai_1999}; \citen{Mira_2001}).  The balance
of these forces results in most bacteria having highly economized
genomes with only a small
fraction (around 10\%) of noncoding
sequences (\citen{Vellai_1998}; \citen{Mira_2001}).
DNA transfer into the bacterial cell
can occur in three ways: (i) transduction by viruses, (ii) conjugation
by plasmids, and (iii) natural genetic transformation (NGT) by
developing competence, a regulated physiological state in which the
bacterial cell is able to take up DNA fragments released by another
cell (\citen{Avery_1944}; \citen{Solomon_1996}). The genetic elements
responsible for transduction and conjugation primarily survive as
parasites, and are located on viral and plasmid DNA.  The genes
required for competence are, however, located on the bacterial
chromosome, placing NGT under the direct control of the cell. While
all three mechanisms play a role in rare gene transfer events between
bacteria of different species, termed horizontal gene transfer, NGT is
the most significant source of active and frequent
genetic transfer within a species (for a comparative review of the
three processes see \citen{Thomas_2005}).  In bacteria capable of NGT,
alleles typically change more frequently by recombination (e.g. 5-10
fold in
\emph{Streptococcus pneumoniae } and \emph{Neisseria meningitids}) than by
mutation
(\citen{LevinBergstrom_2000}; \citen{Feil_2001}; \citen{Feil_2004}; \citen{Thomas_2005}). It
is this combination of high throughput genetic mixing among members of
the same species and direct cellular control that is responsible for
NGT often being referred to as the bacterial analogue of meiotic sex
in eukaryotes (\citen{Smith_1991}; \citen{Smith_1993}).

The persistence of NGT raises the same question as the prevalence of
meiotic sex
(\citen{Bernstein_1985}; \citen{Elena_1997}; \citen{Butlin_2002}; 
\citen{Otto_2002}): What is the short-term advantage of genetic mixing 
to the individual?  NGT is obviously costly, not just because the
machinery of DNA uptake must be maintained, but also because bacteria
undergoing transformation face the risk of incorporating defective
alleles (\citen{Redfield_1997}).  And while the
long-term implications of NGT - for genome adaptation
(\citen{Cohen_2005}) and diversification
(\citen{Holmes_1999}; \citen{Vestigian_2005}) - are clear, the
short-term advantage to individuals (an advantage necessary to
maintain NGT) remains elusive (\citen{Redfield_2001}).

In this work we aim to investigate the role of NGT in bacteria and the
necessary conditions for its short-term maintenance. The ability to uptake naked DNA through NGT
has been detected across a wide phylogenetic spectrum,
ranging from archaea through divergent subdivisions of bacteria, including
representetives from Gram positive bacteria, 
cyanobacteria, {\em Thermus} spp., green
sulphur bacteria and many other Gram negative bacteria 
(\citen{Thomas_2005}).   The details of
transformation, however,  vary widely among bacteria of different
species. With the exception of 
\emph{Neisseria gonorrhoeae} most naturally transformable 
bacteria develop time-limited competence in response to specific
environmental conditions such as altered growth conditions, nutrient
access, cell density (by quorum sensing) or starvation.
The conserved ability among a wide range of bacteria to 
acquire DNA molecules through NGT indicates that the 
genetic trait is functionally important in the environment, 
enabling access to DNA as a source of nutrients or genetic
information (\citen{Thomas_2005}). 

So far it has
been convincingly demonstrated that NGT's role in repairing deleterious
mutations under constant selection is
insufficient for its survival (\citen{Redfield_1997}).  
The lack of other viable explanations has left no alternative except that the
uptake of DNA provides nucleotides for food
(\citen{Redfield_2001}).  This, however, is difficult to reconcile with
the facts that one of the strands is taken up intact (despite the apparent risks of degradation inside the cell), and
that highly specific sequences are required for the binding and uptake
of DNA in some bacterial species, e.g. \emph{N. gonorrhoeae}
or \emph{Haemophilus influenzae} (\citen{Solomon_1996}) (even though nucleotides from other sources confer the same
nutritional benefits). 
In order to understand the relevance 
of NGT as a vehicle of genetic information one must take into
account not only a single population, but a collection of populations
living under diverse and constantly changing ecological conditions, as
only these together possess the complete set of genes common to the
species (\citen{Woese}; \citen{Ochman_2000}). To find the short-term
advantage that maintains NGT, we have to consider its role in allowing
genetic mixing between populations facing variable selection.

\renewcommand\figurename{\textsc{Fig.}}
\begin{figure}[!t]
\centerline{\includegraphics[scale=0.5]{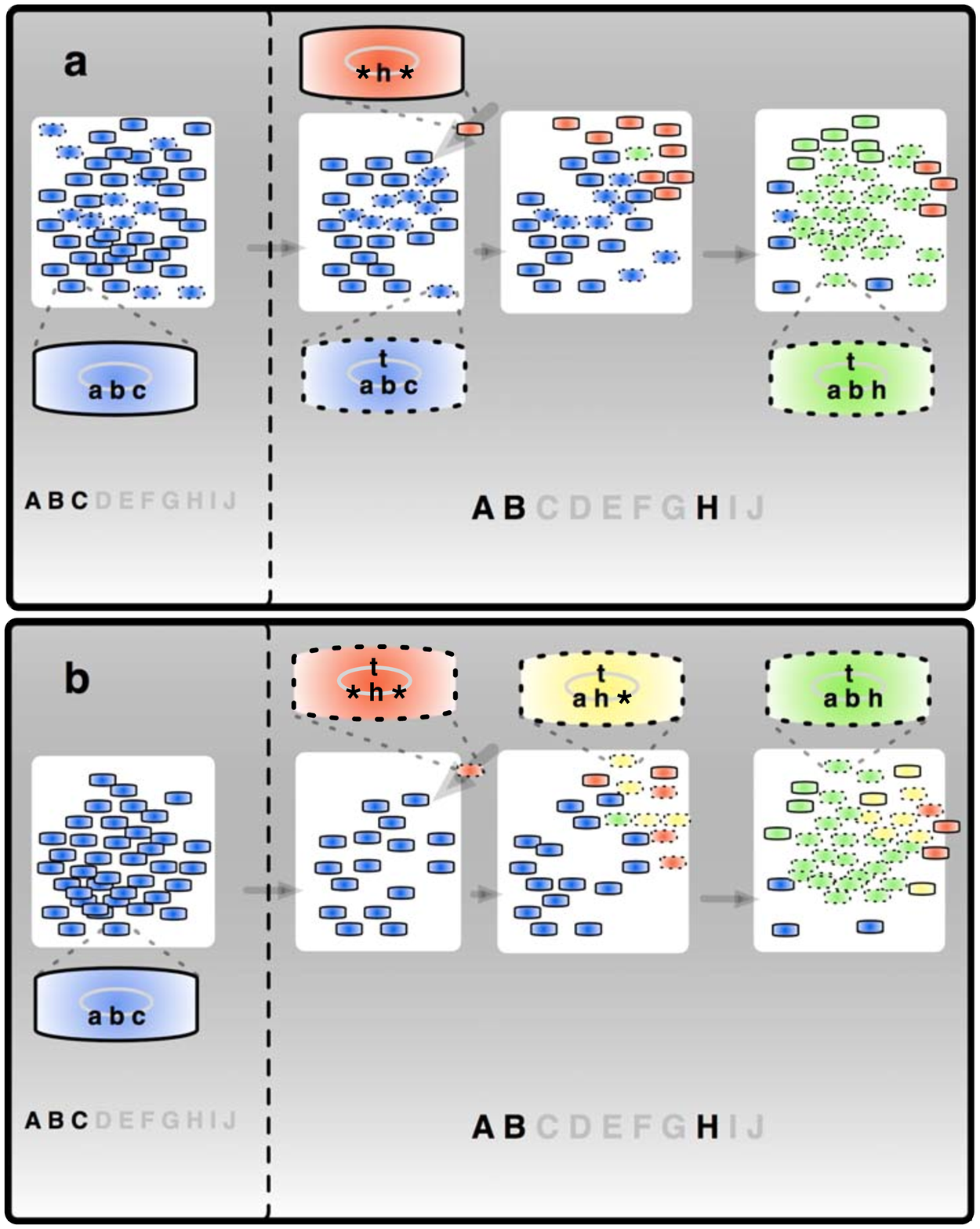}}
\caption{
\setlength{\baselineskip}{14pt} 
{Schematic depiction of the two main processes
    responsible for the spread of NGT.} In both scenarios local food
  types change from \textbf{ABC} to \textbf{ABH}. Bacteria with the
  most viable genotypes are shown with the relevant model genes indicated
  inside, those capable of NGT are emphasized by ``perforated''
  borders.  \emph{(a)} Under rapidly changing conditions the number of
  bacteria with the \textbf{t} (transformation) model gene usually remains significant
  (cf.\ Fig.\ 2b). Newly arrived bacteria possessing the model food gene
  \textbf{h} (red cells) spread due to a lack of competition, subsequently
  allowing local bacteria capable of NGT (blue cells with perforation) to
  successfully adapt by assembling the optimal combination of model genes
  \textbf{abh+t} (green cells with perforation), and to outcompete
  all the others, while also spreading the \textbf{t} model gene. \emph{(b)}
  Under slowly changing conditions, gene \textbf{t} often disappears
  from the local population (cf.\ Fig.\ 2b).  It is, however, possible
  that a competent bacterium possessing gene \textbf{h} (red cells with
  perforation) arrives from another population. Progeny of this
  migrant subsequently adapt and outgrow the local bacteria by
  incrementally (indicated by yellow then green coloring) assembling the best adapted genome,
  \textbf{abh+t} (green cells with perforation), which also results in the spread
  of the model \textbf{t} gene.  }
\label{cartoon}
\end{figure}
\noindent 

\begin{center}
{\large MODEL}
\end{center}

\textbf{The Dynamics of Genome Organization in Bacteria: }
We view a bacterial species as a \emph{metapopulation} (\citen{Hanski}) that is
composed of a large number of spatially distinct populations living
under varying ecological conditions. Different populations experience
selection for different combinations of numerous possible
environmental factors (availability of particular metabolites, host
recognition (\citen{Claverys_2000}) and others
(\citen{Nakamura_2004})).  Most of these factors fluctuate in time
over a broad range of timescales (starting from daily weather changes,
through seasonal alternations and decades-long host life-cycles, up to
long-scale climate changes, to mention just a few).  The populations
are, on one hand, constantly adapting to their own locally changing
environments and, on the other hand, connected by weak
inter-population migration that can span large distances (even
continents and oceans). 
What we aim to show is that under these conditions the advantage of NGT
may lie in providing locally adapted populations with the ability to
respond to environmental changes by importing genes from the
collective gene pool of the species through taking up DNA from
migrants. This way bacteria can economize their genomes by disposing
of genes that are not currently in use in the local environment and
picking up those that have just become useful. There is a growing body
of experimental evidence showing that such genetic plasticity plays a
central role in the adaptation of bacteria, the most well studied
examples being virulence related genetic diversity (particularly of
the genes responsible for capsule composition) in \emph{S.  pneumonae}
(\citen{Claverys_2000}) and the hypervariable region of
\emph{Helicobacter pylori} (\citen{Alm_1999}) responsible for
different pathophysiologies associated with chronic \emph{H. pylori}
infection in humans.

\textbf{Modeling a Bacterial Species: }
To test our hypothesis quantitatively, we consider a model of a
bacterial species (Fig.\ 1) that is capable of living on 10 different
types of food (denoted by \textbf{A}, \textbf{B}, ..., \textbf{J}).
Individual bacteria can utilize any of these foods only if they have
the corresponding model food gene (\textbf{a}, \textbf{b},
..., \textbf{j}, respectively) present in their genomes.  Each such
model gene is understood to represents the complete group of genes
necessary for the utilization of a particular resource type.  While
under natural conditions there may exist dozens of such
resource---gene group pairs, numerical treatment of the model rapidly
becomes intractable as this number is increased, forcing us to
consider only a limited number of food types and gene groups. The
relatively small model genomes used in our simulations (typically a
few model genes long) intend to represent the much larger genomes of
real bacteria. As a consequence, model gene numbers and model genome lengths have to
be suitably rescaled when interpreting the results of the model.

At any moment only 3 out of the 10 possible food sources are
available, but their types are changing in time independently in each
population of the metapopulation. This food change is characterized by
the common rate $R_{\rm food}$, at which one of the food sources is
randomly replaced by another one not currently available in the
population (e.g.,
\textbf{ABC} changes to \textbf{ABH}, and then to \textbf{ADH}, etc.).
Bacteria reproduce at a rate $r_{\rm r}$ (maximized at $r_{\rm r}^{\rm
max}=2$~hour$^{-1}$) that depends on the amount of available food they
can utilize and decreases with the number of functional model food
genes possessed (in order to impose genome economization), for details
see the \emph{Methods} section. Bacteria can also be washed out of their populations at a fixed rate
$r_{\rm w}=10^{-2}$~hour$^{-1}$, lose any of their functional genes by
mutation at a rate $r_{\rm m}=10^{-4}$~hour$^{-1}$ per model gene, and
those possessing a functional copy of the model gene for NGT (denoted
by \textbf{t}) attempt to incorporate exogenous DNA into their genome
from the surrounding medium at a rate $r_{\rm t}=10^{-3}$~hour$^{-1}$
per model gene.  Transformation and mutation are considered at the
level of model genes (\textbf{a} - \textbf{j} and \textbf{t}). That is
to say a single mutation event in the context of our model encompasses
a series of events starting with a deleterious mutation (either a
point mutation or a deletion) leading to loss of function and
continuing with the subsequent gradual loss of the gene group
responsible for the specific function through repeated deletion
events. In other words, we only consider a given group of
genes to be either present and fully functional or completely absent.
We model transformation in a similar fashion, a bacterium may
acquire or lose a complete functioning copy of any model gene
as a result of a single transformation event.
Note that the gene group responsible for transformation (the model
\textbf{t} gene) is capable of eliminating even itself by taking up a
defective copy from the environment.
While the rates of a series
of events leading to the acquisition or loss of a complete gene group
(a single model gene) responsible for a given function are probably
orders of magnitude smaller than those considered in our model, we
argue in detail below (see the \emph{Results} section) that this does
not effect the validity of the results obtained.

The frequency of model gene fragments in the surroundings is
approximated by that in the living individuals of the population
(\citen{Cohen_2005}). This is consistent with the assumption that the
death of bacteria is largely independent of their gene composition,
and experiments showing that DNA fragments from lysed bacteria persist
for hours to days -- as measured in natural transformation
assays (\citen{Thomas_2005}) -- a time-scale that is too short to allow
extracellular DNA to survive the time period between food changes in
our model.  For details of how the competition of individual bacteria
with each possible genotype was modeled see the
\emph{Methods} section.

Migration between the populations is a crucial element of the model,
because without it NGT would just futilely reshuffle the existing
genes inside the populations. To see this, let us suppose that in a
fraction $P$ of a population a certain gene has become
defective. Then the frequency of repairing this
defective gene by taking up a functional copy from a
recently deceased member of the population, $r_{\rm t} P(1-P)$, is the
same as that of accidentally replacing a functional gene by
a defective one, $r_{\rm t} (1-P)P$.  Thus, lacking any
short-term advantage, NGT would rapidly disappear, in agreement with
the results of \citen{Redfield_1997}.  If, however, we take into
account migration, it is able to facilitate the spread of
gene \textbf{t}, as illustrated in the context of our model in Fig.\
1.

 \textbf{Self-Consistent Migration: } Assuming that bacteria can
migrate long distances (such that within this distance a large number
of populations exist with independently changing food sources),
migration can be taken into account very efficiently in terms of
a \emph{mean field} approach, commonly used in statistical
physics. This means that the genotype distribution averaged over the
populations within the migrational range can be well approximated by
the time average of the genotype distribution of any single
population.  In short, spatial and temporal averages are
interchangeable.  Consequently, it is enough to consider only one
population, and use its own history to compute the genotype
distribution of the arriving migrants.  Coupling back, through
migration, an ever increasing fraction of a single population's past
into its own dynamics results in the convergence of the genotype
distribution of the metapopulation to its stationery value in
a \emph{self-consistent} manner, as described in more detail in the
following section. The influx of migrants (the number of incoming
bacteria per unit time) is defined in the model as the product of the
migration rate $R_{\rm migr}$ and the average size of the population
$N$ (which was in the order of $10^8$ in our simulations).

\begin{center}
{\large METHODS}
\end{center}

\textbf{Population Dynamics: } Our population dynamics simulations
were carried out in a manner that allowed the separate treatment of
every bacteria possessing any of the possible model genotypes. The
frequency of individual bacteria of each genotype was calculated by
solving the $2^{10+1}$ differential equations describing the number of
bacteria $n(G)$ in each of the $2^{10+1}$ genome states $G$:
\begin{eqnarray}
\frac{{\rm d} n(G)}{{\rm d} t} &=&
r_{\rm r}(G,\{F\},\{n\}) n(G) - r_{\rm w} n(G)
\nonumber \\ &&
- r_{\rm m} (l_f(G)+\delta_{{\rm{\bf t}},G}) n(G) + {\rm [mut.in]}
\nonumber \\ &&
- r_{\rm t} (10+1) \delta_{{\rm{\bf t}},G} n(G) + {\rm [trf.in]} +
{\rm [migr.in]},
\nonumber
\end{eqnarray}
where $l_f(G)$ is the number of functional model food genes in genotype
$G$, $\delta_{{\rm{\bf x}},G}$ is equal to unity if $G$ contains an
intact model gene {\bf x} and zero otherwise. $n(G)$ is treated as a
continuous variable, but with a lower cutoff at $n(G)=1$ to mimic the
discrete nature of bacterial populations. This lower cutoff is
implemented for each subpopulation with $n(G)<1$, by resetting 
 $n(G)$ to $1$ with probability $n(G)$, or to $0$ with probability
$1-n(G)$, at each time step of the simulation.
 The first term on the right
hand side describes the reproduction of bacteria as detailed
below. The second term corresponds to the washout of bacteria from the
population. The third and forth terms deal with mutation, the former
one considers bacteria with genotype $G$ that undergo mutation, while
the later one ${\rm [mut.in]}$ is the sum of contributions from all
bacteria that mutate into state $G$.  The fifth and sixth terms
describe transformation, the former one corresponding to those
bacteria in state $G$ that undergo transformation, while the later one
${\rm [trf.in]}$ represents the complicated sum of all transformations
that lead to state $G$.  The last term ${\rm [migr.in]}$ corresponding
to migration is also described below.

\renewcommand\figurename{\textsc{Fig.}}
\begin{figure}[!t]
\centerline{\includegraphics[scale=0.3]{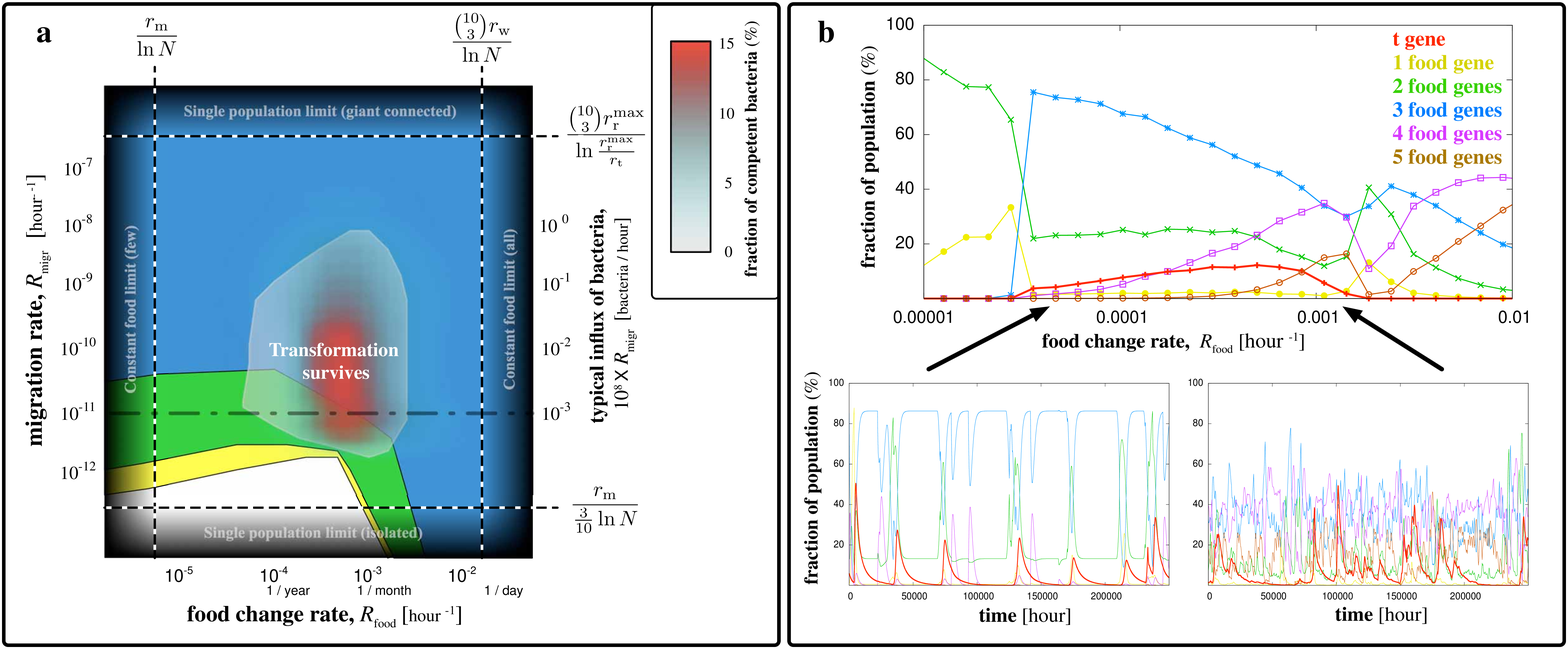}}
\caption{ \setlength{\baselineskip}{14pt}
Phase diagram for the persistence of NGT.
 \emph{(a)} The average percentage of bacteria capable of NGT is
  shown with color gradient (translucent from white to red) on the $R_{\rm food}$ - $R_{\rm
   migr}$ parameter space, with different points
    corresponding to different possible bacterial metapopulations.
  The maximum genome size (the number of functional
  model food genes) in systems where transformation has been turned off is underlain in color code (yellow 
  regions only have bacteria with one functional food gene,
  green regions have bacteria with at most two,
  while the blue region corresponds to metapopultions where
  bacteria exist that can utilize all three food types). 
The dashed lines indicate the
  theoretical limits of the persistence of transformation as described in the text.  
  \emph{(b)} The upper panel shows a slice in the parameter space
  along the dashed-dotted black
  line in part \emph{(a)}, with data points corresponding to the fraction of bacteria
  with the {\bf t} gene and the fraction of those having exactly
  $1,2,3,\dots$ functional food genes. The lower panel displays two
  illustrative time series from the simulations, corresponding to a low 
  and a high  food change rate, as indicated by the arrows.}
\label{map}
\end{figure}

\textbf{Reproduction Rate: } The fitness of individual bacteria with a
given genotype $G$ is influenced both by available food
$\{F\}\equiv{F_1,F_2,F_3}$, as well as genome length
$l(G)=l_f(G)+l_{\rm t}\delta_{{\rm{\bf t}},G}$, where $l_t$ denotes
  the length of the {\bf t } gene relative to the model food genes
  (and is considered to be less than unity, as explained below). Available food types $F_i$
were each considered to be constantly replenished in the environment,
each type being characterized by a strength $S_i$ corresponding to the
number of bacteria the given food type could sustain at the maximum
division rate $r_{\rm r}^{\rm max}$.  Each bacterium with a functional
copy of model gene $f_i$ (corresponding to a given food type $F_i$)
present in its genome ($\delta_{f_i,G} = 1$) received an ``equal
share'' $S_i/ \sum\limits_{G'} \delta_{f_i,G'} n(G')$ of this food,
while those bacteria that did not have an intact copy of the model
gene ($\delta_{f_i,G} = 0$) in their genome received none.  In general
the speed at which a given bacterium can divide $r_{\rm
r}(G,\{F\},\{n\})$ is proportional to the total amount of food it can
utilize $\sum_{i=1}^{3} \delta_{f_i,G}
S_i/ \sum\limits_{G'} \delta_{f_i,G'} n(G')$, but also decreases by a
factor of $2 / (1+l(G)/l_{\rm opt}) $ as the total number of intact
model food genes present in its genome increases, where $l_{\rm
opt}=3$ is the optimal genome size.  The denominator of the latter
factor takes into account that the time necessary for a bacterium to
divide has a component proportional to genome length.
Bacteria that posses only those three model genes
that are necessary for the utilization of the three food types
available at a given moment are the ones that divide the fastest
(i.e., have the highest fitness), their genomes being the most highly
economized. As outlined in the \emph{Model} section real bacteria may posses
several dozen gene groups necessary for the utilization of fluctuating
resources, but as we are not able to treat numerically more than a few
model gene---food pairs, we have chosen the parameters of the fitness
function such that it decreases rapidly as the number of model genes
grow (e.g.\ by 33\% if twice the optimal number of model genes is
present),  that is we, in this sense, consider each model gene to correspond to several
gene groups.  Further, as the gene group responsible for DNA uptake
(the \textbf{t} model gene) is only one of several dozens
 and in
reality consist
of a relatively small number of genes, we have included it with a smaller relative length ($l_{\rm t}=0.1$) 
in the total genome length. 
Due to practical constraints, however,
the mutation rate of the {\bf t} gene was considered
(except for the data presented in Fig.\ 3c.)
to be the same as that of the food genes (which corresponds to a
tenfold increase in the mutation rate for the {\bf t} gene).
Therefore, in our
simulations the reproduction rate of a bacterium with genotype $G$ and
available foods $\{F\}$ was given by:
\begin{eqnarray}
r_{\rm r}(G,\{F\},\{n\}) = 
\frac{2 \enskip r_{\rm r}^{\rm max} }{1+l(G)/l_{\rm opt}}
\min \!\!\left[
1, \sum_{i=1}^{3}
\frac{\delta_{f_i,G} S_i}{\sum\limits_{G'} \delta_{f_i,G'} n(G')}
\right]\!\!, 
\nonumber
\end{eqnarray} 
 and we chose the food strength to be $S_i=10^5$ for each available
 food type.

\textbf{Migration: }
Since all food types are equivalent and consequently all combinations
of these are equally likely, the number of mean-field variables needed
to describe the global genotype distribution can be reduced from
$2^{10+1}$ to $10 \times 2$ corresponding to all genotypes $G$ with
$1,2,\dots,10$ intact metabolic genes, with and without the
$\mathrm{\bf t}$ gene. To take into account the influx of bacteria
from external populations we took the averages of these $10\times 2$
types in the population with a sliding-growing time window (always
encompassing the last quarter of the simulation) and subsequently
calculated the migrational term ${\rm [migr.in]}$ for each genotype
$G$ by multiplying the corresponding mean-field variable with the
appropriate combinatorial factor and the migration rate $R_{\rm
migr}$.

\begin{center}
{\large RESULTS}
\end{center}

\textbf{Phase Diagram for the survival NGT: }
Performing extensive computer simulations for various values of the
two main external parameters, the food change rate $R_{\rm food}$ and
the migration rate $R_{\rm migr}$, we have found that NGT (represented
by model gene \textbf{t}) indeed survives under a wide range of
parameters, as can be seen in Fig.\ 2a. In these stochastic population
dynamics simulations we have numerically followed the time evolution
of the number of bacteria in each of the possible $2^{10+1}$ genotypes
(representing the presence or absence of the 10 model food genes and
the model transformation gene) in a single population, as detailed in
the
\textit{Methods} section.

{\bf \textbf{Limits for the Persistence of NGT}}

The limits beyond which NGT cannot persist (dashed straight lines in
Fig.\ 2a), either because the temporal or the spatial fluctuations
become irrelevant, can be estimated easily. If $R_{\rm food} < r_{\rm
m} / \ln N$, then the food sources remain unchanged for such a long
time that the \textbf{t} gene completely disappears by deletion before
it could become beneficial at the next food change.  At the other
extreme, for $R_{\rm food} > {10 \choose 3} r_{\rm w} / \ln N$, i.e.,
when the rate at which any given food combination recurs ($R_{\rm
food} / {10
\choose 3}$) is larger than the rate at which bacteria carrying the
corresponding combination of functional genes are completely
washed out of the population (max.\ $r_{\rm w} / \ln N$),
transformation cannot confer an advantage through assembling this
combination as it is always present in the local population. In other
words, the food changes so rapidly that populations effectively
experience a constant feeding (with all possible food combinations),
and NGT becomes useless.

There are similar constraints on the migration rate as well.
Obviously, for very small migration rates, $R_{\rm migr} N < r_{\rm m}
/ (\frac{3}{10} \ln N$), i.e., when the influx of bacteria with a
newly required food gene ($\frac{3}{10} R_{\rm migr} N$) is lower than
the rate of the complete disappearance of the functional \textbf{t}
gene ($r_{\rm m} / \ln N$), migration loses its role in the
propagation of NGT, thus, the metapopulation practically falls apart
into isolated populations, in which transformation cannot survive.
The other limit is $R_{\rm migr} N > {10 \choose 3} r_{\rm r}^{\rm
max} /
\ln \frac{r_{\rm r}^{\rm max}}{r_{\rm t}}$, 
i.e., when the migration is so intense that after a food change the
rate of the arrival of a bacterium with the optimal combination of
functional model genes ($R_{\rm migr} N / {10
\choose 3}$) exceeds the rate at which the model gene that has just become
beneficial proliferates and subsequently gets incorporated by a member
of the original population ($r_{\rm r}^{\rm max} / \ln \frac{r_{\rm
r}^{\rm max}}{r_{\rm t}}$). Then the entire metapopulation effectively
becomes a single giant population, in which transformation cannot
survive either. Although these are rather crude estimates, which only give the limits
 outside of which transformation is certain not to survive, we have
 found none the less, that within these extremes
 NGT persists for most parameter values, indicating the robustness of
 our proposed mechanism.

\textbf{Effects of NGT on Genome Organization: }
To demonstrate the dramatic effect of transformation on the genome
composition, we take a cut through the parameter space (dashed-dotted
line in Fig.\ 2a) and plot the average fraction of bacteria possessing
a functional copy of the \textbf{t} gene, as well as the fraction of those having exactly 1, 2, ...
functional model food genes in Fig.\ 2b. For low food change rate, but
within the range where NGT survives, most bacteria only contain
functional copies of the three necessary food genes, and the average
fraction of functional copies of the model gene \textbf{t} is very low
as it is used very rarely.  Functional copies of the \textbf{t} gene
can occasionally disappear from the population and then subsequently
be replanted by migrants as illustrated in Fig.\ 1a. For larger food
change rates, but still within the survival range of NGT,
the \textbf{t} gene becomes beneficial more often, its average
fraction increases (except near the other end of the range), and the
fraction of bacteria having four or more functional model food genes
also increases. Two time series (at a low and a high food change rate)
displaying the evolution of these fractions are shown in the lower
panel of Fig.\ 2b.

\renewcommand\figurename{\textsc{Fig.}}
\begin{figure}[!t]
\centerline{\includegraphics[scale=1.]{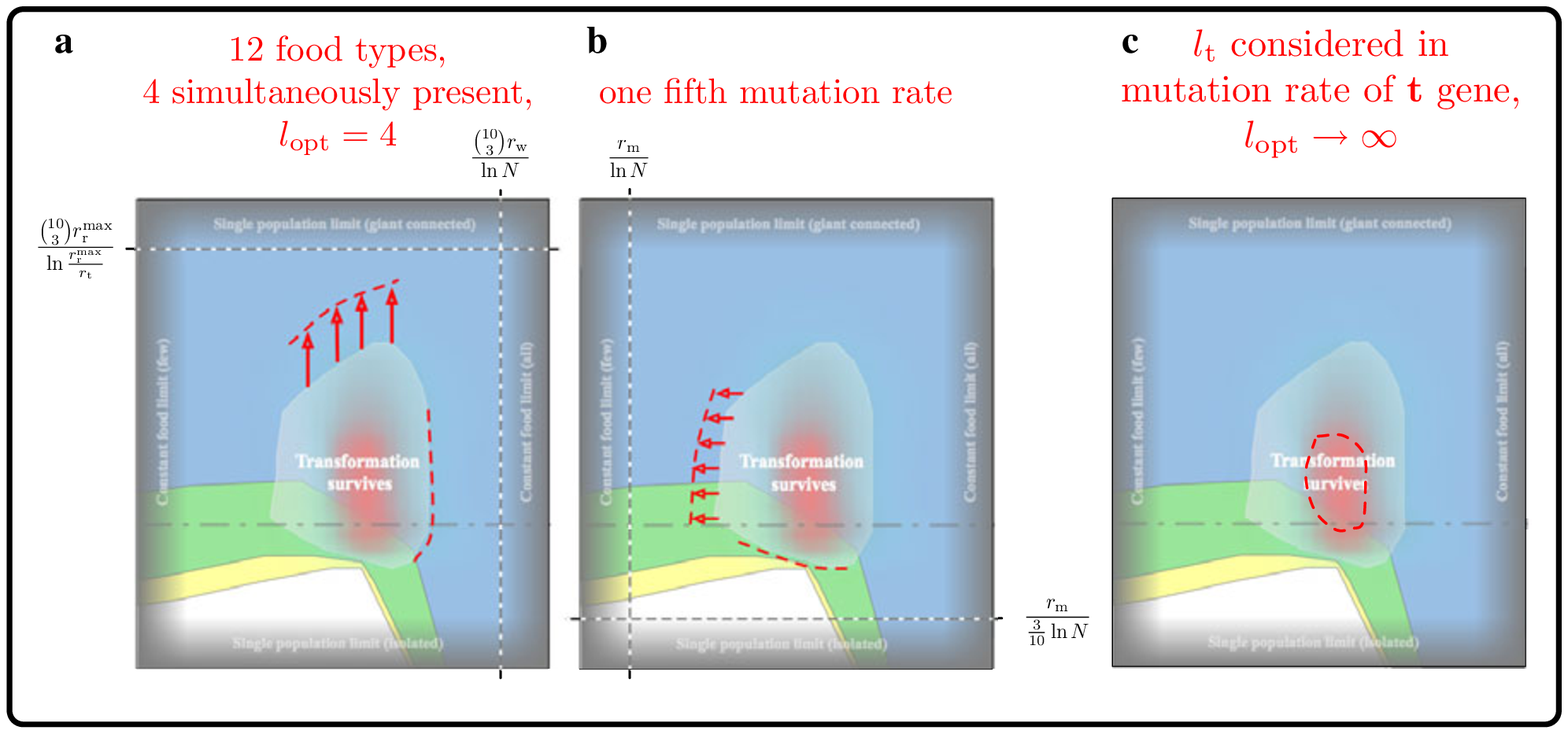}}
\caption{ \setlength{\baselineskip}{14pt}
Robustness of the proposed mechanism for the survival of NGT.
We investigated the robustness of the the survival range of NGT by
varying several model parameters: In {\em (a)} we changed the number
of possible food types from 10 to 12 and the number simultaneously present
from 3 to 4, consequently also setting $l_{\rm opt}=4$. The survival range of
NGT was found to extend to higher values of the migrational influx in
accordance with our predictions, while it only extended to a much
smaller degree toward faster food change rates. In {\em (b)} we have changed the mutation
rate to $1/5$ of that considered in Fig.\ 2. Under these conditions
--similarly to {\em(a)}--
the survival range of NGT extends in one direction -- toward smaller food change rates,
but does not significantly move in the other -- toward smaller values of migrational
influx. In the third panel {\em (c)}
we demonstrate that the {\bf t} gene survives, although under a more
limited set of parameters, even if selection for genome length is
absent. During these simulations we gradually approached the limit $l_{\rm opt} \rightarrow \infty$, where
selection for genome length disappears. We also increased the
mutation rate of food genes by an order of magnitude, while keeping
that of the {\bf t} gene fixed. This way we were able to take into
consideration,
the length of the {\bf t} gene $l_{\rm t}=0.1$ in terms of the
relative mutation, while also keeping our simulations tractable.}
\label{robustness}
\end{figure}

\textbf{Survival Range of NGT: }
In the model we have chosen numerically tractable values
for the rates ($r_{\rm r}^{\rm max}$, $r_{\rm w}$, $r_{\rm m}$,
$r_{\rm t}$) characterizing a given bacterial species.
Some of these rates may be off by up to a few orders of magnitude for
certain types of bacteria, this, however, should not fundamentally effect the width of the
parameter range where NGT survives (neither in terms of $R_{\rm food}$
or $R_{\rm migr}$), as this primarily depends on the binomial factor
${10 \choose 3}$.  Moreover, under natural conditions the number of
possible food types is usually much larger than 10, and the number of
simultaneously available ones is also larger than 3, thus the survival
range of NGT is probably even broader (extending to higher values of
$R_{\rm food}$ and $R_{\rm migr}$) than what our calculations predict.
We have attempted to survey the effect of more realistic, but 
computationally more demanding parameter sets as presented in Fig.\ 3a
and 3b, and found that the survival range of NGT indeed became broader.
In Fig.\ 3a we increased the number of possible food types from $10$ to
$12$, while also increasing the number present at any on time from $3$ to
$4$, consequently also setting $l_{\rm opt}=4$. In Fig 3.b we have
decreased the mutation rate to one fifth of the value considered
previously considered. 

Characterizing transformation by a single rate coefficient $r_{\rm t}$
is also a rather crude simplification, since many bacteria become
competent only under certain conditions. Fortunately, the limits for
the survival of NGT calculated above depend very weakly on the value
of $r_{\rm t}$. Besides, the parameter range where a better regulated
competence will persist is expected to be even larger than it is in
our simplified model.

Up to now we have only considered food sources that fluctuate with a
single characteristic rate $R_{\rm food}$.  To check what happens in
more complex situations, we have added an equal amount of constant
food source (\textbf{XYZ}) to the fluctuating ones. These simulations
have confirmed that the survival range of NGT remains virtually
unaffected, indicating that the ability to lose and reload a few types
of intermittently beneficial ``operational'' genes is advantageous to
the bacteria and sufficient to maintain NGT.

Migration between bacterial populations in nature clearly depends on
distance and, as a consequence, a wide variety of migration rates are
usually present in the metapopulation. Populations close to each other
(with strongly correlated environmental fluctuations or with intense
inter-population migration), however, may be grouped together and
considered as a large effective population. For NGT to persist it is
sufficient that a reasonable number of such ``effective'' populations
exist in the metapopulation, a requirement that does not seem
unrealistic in face of the highly varied conditions under which
bacteria prevail on Earth.

Finally one very important question remains which we have not addressed
so far: Is selection for shorter genome length necessary for the survival of NGT
as a mechanism to reload genes?  To answer this question we preformed
simulations where we gradually approached the limit 
$l_{\rm opt} \rightarrow \infty$, where selection for genome length disappears. 
We found that if we took into consideration the length of the {\bf t}
gene $l_{\rm t}$ in terms
of the mutation rate, the {\bf t } gene survived (see Fig.\ 3c). Because,
smaller mutation rates imply longer convergence times in our
simulation, we implemented the value of $l_{\rm t}=0.1$, not by
decreasing the mutation rate of the {\bf t} gene, but by increasing
those of the food genes by a factor of ten. From these results we can conclude that while selection for
shorter genome length substantially increases the size of the parameter
range where NGT is able to survive solely as a mechanism to reload genes, it is not indispensable.
We may further argue that the relatively small region in Fig.\ 3c
where the {\bf t} gene persist, should
become much larger for more realistic values of the mutation rate and the
number of fluctuating food types (cf. Fig.\ 3a and 3b).
\begin{center}
{\large DISCUSSION}
\end{center}

In light of our simulations, we suggest that 
the existence of NGT is facilitated by
its role as a vehicle to reload genes.  We argue that the short-term
advantage that sustains NGT long enough for its evolutionary effects
to emerge, lies, at least in part, in providing mobility to variably
selected genes. It allows individuals to reload genes lost from a
population -- due to long disuse -- but still available in the
metapopulation, bringing together genes from the collective gene pool
of the species with locally adapted genomes.  This advantage prevails
if spatio-temporal fluctuations (\citen{Meyers_2002}) in the
environment (imposing variable selection pressure on different
populations of the same species) exist in parallel with weak
 migration between the populations (allowing genetic mixing). 

Whether or not natural bacterial populations actually experience
the kind of population subdivision and inter-population
migration necessary for our model to be applicable remains to be
demonstrated experimentally. Some examples which may easily fulfill
these conditions, however, readily come to mind, e.g.\ experience shows that any
perishable substance that is a potential food source for
bacteria is promptly colonized; one may also consider the intestinal
flora of grazing animals, herds of which cover large distances while
occasionally encountering each other at locations, such as water
sources, where migration may occur between the 
microbial populations resident in their intestines.

Provided that the above conditions may be
rather general, our results compel us to imply that the ability 
of active DNA uptake may easily
have evolved through the gradual specialization of a more general
transport mechanism (\citen{Woese}), driven not by the need to serve
the cell with nucleotides for food, but by an advantage conferred
through providing homologous sequences for restoring genes eroded by formerly adaptive
deletions. In other words NGT is not an accidental byproduct of
nutrient uptake, but has come into existence in order to
counterbalance gene loss, which inevitably occurs in highly economized
genomes under fluctuating selection pressure. 
It should be emphasized, however, that this does not
  exclude the advantage transformation confers through enabling 
access to DNA as a source of nutrients, which  most probably helps sustain NGT. 

A fundamental long-term
effect of genetic mixing by NGT, with important implications for the
understanding of the evolution of eukaryotic sex
(\citen{Holmes_1999}; \citen{Smith_EvSex};
\citen{Szathmary_Book};), is that it prevents bacterial species from falling
apart into independently evolving clonal lineages
(\citen{Dykhuzien_1991}), by facilitating genetic mixing between
genomes of sufficient homology.  From a medical perspective the
dynamic nature of bacterial genomes has clear significance for the
pressing problem of the rapid spread of antibiotic resistance and
other pathogenic traits (\citen{Claverys_2000}).
\vskip 0.5cm
I. D. acknowledges support from
the Hungarian Science Foundation (Grant No. OTKA K60665).
\newpage
\begin{center}
{\large LITERATURE CITED}
\end{center}

\end{document}